# KrF pulsed laser deposition of chromium oxide thin films from $Cr_8O_{21}$ targets

N. POPOVICI[1], M.L. PARAMÊS[1], R.C. DA SILVA[2], O. MONNEREAU[3], P.M. SOUSA[1],

A.J. SILVESTRE[4] AND O. CONDE[1a]

[1]Dept. Física, Faculdade de Ciências da Universidade de Lisboa, Campo Grande, Ed. C8,

1749-016 Lisboa, Portugal

[2]Dept. Física, Instituto Tecnológico e Nuclear, E.N. 10, 2685 Sacavém, Portugal

[3]Université de Provence, CNRS UMR 6121, Centre St. Jérôme, 13397 Marseille Cedex 20, France

[4]Instituto Superior de Engenharia de Lisboa, R. Cons. Emídio Navarro, 1749-014 Lisboa, Portugal

**ABSTRACT**   Chromium oxides, $Cr_xO_y$, are of great interest due to the wide variety of their technological applications. Among them, $CrO_2$ has been extensively investigated in recent years because it is an attractive compound to be used in spintronic heterostructures. However, its synthesis at low temperatures has been a difficult task due to the metastable nature of this oxide. This is indeed essential to ensure interface quality and the ability to coat thermal-sensitive materials such as those envisaged in spintronic devices. Pulsed Laser Deposition (PLD) is a technique that has the potential to meet the requirements stated above.

In this work, we describe our efforts to grow chromium oxide thin films by PLD from $Cr_8O_{21}$ targets, using a KrF excimer laser. The as-deposited films were investigated by X-ray diffraction and Rutherford backscattering spectrometry. Structural and chemical composition studies showed that the films consist of a mixture of amorphous chromium oxides exhibiting different stoichiometries depending on the processing parameters, where nanocrystals of mainly $Cr_2O_3$ are dispersed. The analyses do not exclude the possibility of co-deposition of $Cr_2O_3$ and a low fraction of $CrO_2$.

**PACS:** 81.15.Fg, 75.50.Dd

## 1. Introduction

Chromium oxide thin films are of great interest due to the wide variety of their technological applications. The most stable phase is the corundum structured $Cr_2O_3$. This form of oxide has important industrial applications, for instance in catalysis [1] and solar thermal energy collectors [2]. Currently, low-reflective $Cr_2O_3/Cr$ films are widely used as black matrix films in liquid crystal displays [2]. As this chromium oxide is an insulating antiferromagnetic material it is also suitable as a tunnel junction barrier [3].

Chromium dioxide ($CrO_2$) has recently attracted considerable attention because it is strongly ferromagnetic at room temperature ($T_C = 393$ K) and has a half-metallic band structure fully spin-polarised at the Fermi level [4-7], making it attractive for use in

---

[1a] Corresponding author: Tel: +351 217500035, Fax: +351 217500977, email: omconde@fc.ul.pt

spintraonic heterostructures. Therefore, much effort has been put into developing efficient and controlled methods for preparing $CrO_2$ films at sufficiently low temperatures. Nevertheless, the synthesis of $CrO_2$ films has been a difficult task due to its metastable nature – $CrO_2$ is metastable at atmospheric pressure and, if heated, easily decomposes into the insulating antiferromagnetic $Cr_2O_3$ phase.

Epitaxial thin films of $CrO_2$ have been grown on $TiO_2$ (100) and $Al_2O_3$ (0001) by thermal CVD using $CrO_3$ as chromium precursor [8-13]. Pulsed laser deposition (PLD) has also been successful in growing $CrO_2$ on Si(111) and $LaAlO_3$ [14] from a $Cr_2O_3$ target; however, a high substrate temperature of 390 ºC was also required in the later technique. Despite these achievements, there is a continuous search for deposition methods that allow the growth of $CrO_2$ thin films at low temperature.

In this paper, we present results on chromium oxide films produced onto Si (100) by room temperature PLD in oxygen atmosphere and using $Cr_8O_{21}$ as target material.

## 2. Experimental details

### 2.1 Film growth

Chromium oxide films were grown by reactive pulsed laser deposition (RPLD) using a stainless steel HV deposition chamber and a pulsed UV laser (KrF, 248 nm wavelength, 30 ns pulse duration) with associated beam delivery optics. The laser was operated at 5 Hz and the laser beam was incident at an angle of 45º with respect to the target surface. The target was produced from $Cr_8O_{21}$ powder, as described below, and was continuously rotated and periodically translated during laser ablation to renew the irradiated surface and to prevent crater formation. The ablated material was collected onto a Si(100) substrate, at room temperature, placed in front of and at 6.2 cm from the target. Prior to an experiment, the chamber was evacuated to $2-7 \times 10^{-4}$ Pa, a screen was placed between the substrate and the target for protection, and the target surface was cleaned for several minutes by irradiating with the non-focused KrF laser beam. During deposition, oxygen (99.999 %) was flown through the chamber and the background pressure was varied between 0.3 and 1.9 Pa. The laser fluence was kept at $4.0 \pm 0.3$ J cm$^{-2}$.

## 2.2 Target preparation

Powder of $Cr_8O_{21}$ was obtained by slow thermal decomposition of $CrO_3$ powder in air at 260 ºC. Afterwards, the powder was crushed to nanometre size and sintered at 220 ºC and 0.2 GPa into pellets 20 mm in diameter and 2 mm thick.

## 2.3 Sample characterisation

Structural analysis of the films was carried out in a Siemens D5000 diffractometer by glancing incidence X-ray diffraction (GIXRD), using CuK$\alpha$ radiation at 1º angle of incidence to the specimen surface. The identification of crystalline phases was done using the JCPDS database cards [15].

Chromium distribution profiles were measured by Rutherford Backscattering Spectrometry (RBS). All the RBS analyses were performed in a 3 MV Van de Graaff accelerator, using 2.0 MeV He$^+$ and H$^+$ beams. The backscattered ions were detected by means of two silicon surface barrier detectors, placed at angles of 140º and 180º to the beam direction, and with energy resolutions of 13 and 18 keV, respectively.

## 3. Results and discussion

Glancing incidence XRD patterns from films deposited at RT with different oxygen pressures are depicted in Fig.1. As can be seen, all the diffractograms show narrow diffraction lines superimposed on wide humps indicating that the films consist of crystallites embedded in an amorphous or nanocrystalline matrix. The size of the crystallites was estimated between 45 and 65 nm from XRD data using the Scherrer equation. By comparing the data with JCPDS database, one can conclude that for all the samples these nanocrystals are mainly of $Cr_2O_3$ phase, although other phases such as $Cr_3O_4$ and $Cr_2O_5$ may also be present.

The amorphous region of the diffractograms, related to the matrix, moves to higher $2\theta$ values as oxygen background pressure decreases. For sample O1, deposited with an oxygen pressure of $p_{O2}$ = 1.7 Pa, the hump comprises the 20º – 35º region where both the most intense diffraction lines from $CrO_3$ and the highest concentration of $Cr_8O_{21}$ peaks are present. For sample O4, grown at $p_{O2}$ = 0.5 Pa, the hump extends from 30º to 40º. This region is dominated by the presence of diffraction lines due to $Cr_3O_4$ and $Cr_2O_3$ compounds. Sample O5, deposited with an intermediate oxygen pressure of 1.3 Pa, displays a less intense background indicating that the matrix could be more amorphous than in the two previous samples. In this case, the weak hump is observed for intermediary $2\theta$ values i.e. around 25º to 35º, and the presence of diffraction lines from $Cr_2O_5$ in addition to $Cr_2O_3$ is clearly shown. Therefore, a dependence of the matrix composition on the oxygen background pressure could be established: for lower values of $p_{O2}$ (curve O4) the matrix is based on compounds with low O/Cr ratio such as $Cr_3O_4$ (1.3/1.0) and $Cr_2O_3$ (1.5/1.0). At higher $p_{O2}$ (curve O1), the matrix becomes richer in oxygen and phases such as $CrO_3$, $Cr_8O_{21}$ or $Cr_2O_5$ (2.5 < O/Cr < 3.0) can be used to fit the data. From GIXRD analysis, we speculate that the most promising conditions for preparing chromium oxide thin films containing the ferromagnetic $CrO_2$ oxide should stay close to those used for sample O5. In fact, this sample displays a very uniform dark grey colouration and the presence of a small amount of $CrO_2$ in the background matrix cannot be excluded.

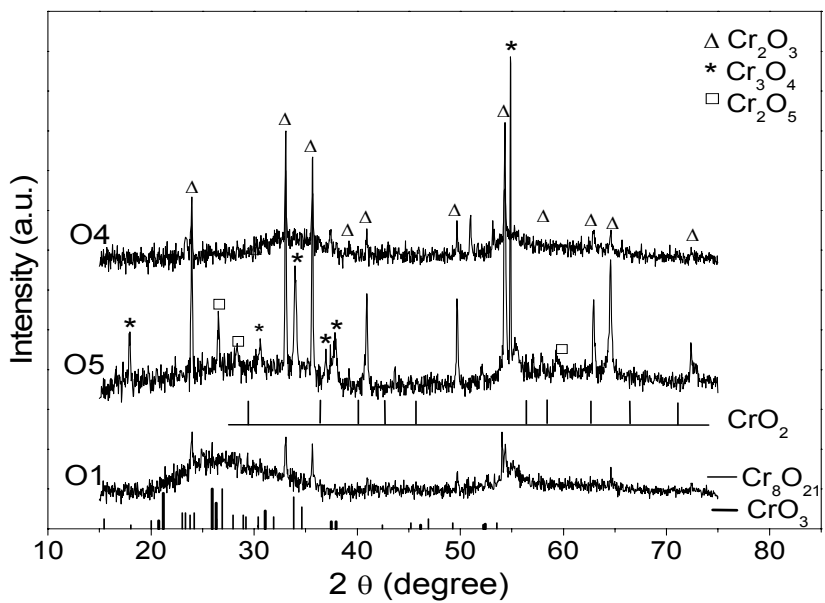

**FIGURE 1** GIXRD patterns of $CrO_x$ samples made at different oxygen background pressures: 1.7 Pa (O1), 1.3 Pa (O5) and 0.5 Pa (O4). Labelled lines and patterns were taken from the JCPDS database cards.

In order to help clarifying the chemical composition of sample O5, a comparative study of this sample and sample O1 was carried out by Rutherford Backscattering. The RBS spectra obtained from the as-deposited films are displayed in Fig. 2.

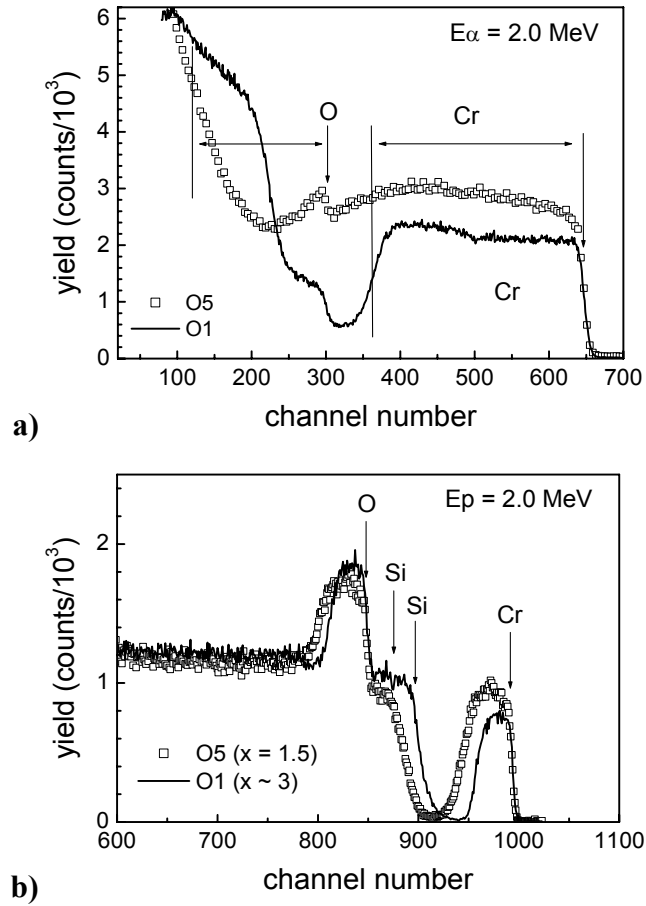

**FIGURE 2** RBS spectra of chromium oxide films grown onto Si (100), recorded with 2.0 MeV He$^+$ (a) and H$^+$ (b) beams.

From these spectra, recorded both with He ion and proton beams, an atomic concentration ratio O/Cr of 3/1 was calculated for sample O1, in good agreement with the X-ray analysis previously presented. The fact that this ratio is constant along the film depth allows to assume that the film is homogeneous with the density of $CrO_3$, $\rho$ = 2.7 g cm$^{-3}$, which leads to a chromium profile extending over a depth of 900 – 1100 nm. This thickness value agrees well with the one measured in cross-section scanning electron micrographs (not shown). A similar analysis was performed for sample O5. The O/Cr atomic concentration ratio deduced was 1.5 consistent with $Cr_2O_3$ formation.

However, it should be noted that RBS does not allow distinguishing among composite oxides where one oxide type is prominent. This is the case if sample O5 would consist of a mixture of e.g. 98% $Cr_2O_3$ + 2% $CrO_2$, 94% $Cr_2O_3$ + 6% $CrO_2$ or 98% $Cr_2O_3$ + 2% $CrO_3$. All these combinations give an O/Cr ratio very close to 1.5 and, therefore, the presence of $CrO_2$ in this sample cannot be excluded on the basis of RBS, as for the X-ray data analysis.

## 4. Conclusions

Ablation of $Cr_8O_{21}$ targets was performed in oxygen background, at various pressures, using a pulsed UV KrF laser. The ablated material was collected onto Si (100) substrates to grow thin films. The chemical composition of the samples was evaluated by Rutherford backscattering with both $He^+$ and proton beams, and their morphology was studied by scanning electron microscopy. X-ray diffraction carried out at glancing incidence angle was used for phase analysis. The films consist of $CrO_x$ oxides whose nature depends strongly on oxygen content in the gas phase, and of nanocrystals, mainly composed of $Cr_2O_3$ which is the most stable phase in the Cr – O phase diagram. For some limited conditions, there is evidence of co-deposition of $Cr_2O_3$ and $CrO_2$, the latter phase being only present in very small fraction.

**ACKNOWLEDGEMENTS** This work was supported by the EU contract FENIKS: G5RD-CT-2001-00535.